\documentclass[%
 reprint,
 amsmath,amssymb,
 aps,
]{revtex4-2}

\usepackage{graphicx}% Include figure files
\usepackage{dcolumn}% Align table columns on decimal point
\usepackage{bm}% bold math
\usepackage{subcaption}
\usepackage{amsmath}
\usepackage{siunitx}

\begin{document}

\preprint{APS/123-QED}

\title{Investigating the signs of evolutionary characteristics in the energy spectrum of shock wave acceleration}% Force line breaks with \\

\author{Xu-Lin Dong}
\affiliation{College of Physics, Hebei Normal University, No. 20 Road East 2nd Ring South, Shijiazhuang, 050024 Hebei, China}
\affiliation{Key Laboratory of Particle Astrophysics, Institute of High Energy Physics, Chinese Academy of Sciences, No. 19 B Yuquan Road, Shijingshan District, Beijing, 100049, China}
\author{Wei-Kang Gao}
\affiliation{Key Laboratory of Particle Astrophysics, Institute of High Energy Physics, Chinese Academy of Sciences, No. 19 B Yuquan Road, Shijingshan District, Beijing, 100049, China}
\affiliation{University of Chinese Academy of Sciences, No. 19 A Yuquan Road, Shijingshan District Beijing 100049, China}

\author{Yi-Qing Guo}\email{guoyq@ihep.ac.cn}
\affiliation{Key Laboratory of Particle Astrophysics, Institute of High Energy Physics, Chinese Academy of Sciences, No. 19 B Yuquan Road, Shijingshan District Beijing 100049, China}
\affiliation{University of Chinese Academy of Sciences, No. 19 A Yuquan Road, Shijingshan District Beijing 100049, China}
\affiliation{TIANFU Cosmic Ray Research Center, No. 1500 Kezhi Road Chengdu, 610101 Sichuan, China}

\author{Shu-Wang Cui}\email{cuisw@hebtu.edu.cn}
\affiliation{College of Physics, Hebei Normal University, No. 20 Road East 2nd Ring South, Shijiazhuang, 050024 Hebei, China}
\affiliation{TIANFU Cosmic Ray Research Center, No. 1500 Kezhi Road Chengdu, 610101 Sichuan, China}

\begin{abstract}
Under ideal conditions, the theory of shock acceleration for cosmic rays predicts that different elements should exhibit strictly identical spectral indices ($\gamma$) when accelerated to the same rigidity (R).
However, recent high-precision measurements of elemental energy spectra  have definitively established the existence of variations in spectral indices ($\gamma$) across different elements. This study constrains the spectral indices ($\gamma$) of cosmic-ray elements using AMS-02 and DAMPE observations within the Spatially Dependent Propagation (SDP) model. For elements with $ {\mathrm{A}}/{\mathrm{Z}} \approx 2$, $\gamma$ shows significant positive correlations with both atomic number Z and mass number A, likely due to A or Z-dependent fragmentation cross-sections. Predictions indicate that the observed spectra of Ni and Zn will align with the Fe spectrum, while their injection spectra will exhibit slightly softer spectral indices  compared to Fe. Future observations from AMS-02, DAMPE and HERD are expected to verify these findings, while theoretical models are needed to systematically explain this phenomenon.

%In an ideal scenario, the theory of shock acceleration of cosmic rays predicts identical spectral indices ($\gamma$) for different elements after acceleration. However, studies of the energy spectra of elements with $\left \langle {\mathrm{A}}/{\mathrm{Z}} \right \rangle \approx 2$ measured by the AMS-02 experiment have confirmed the existence of differences in the spectral indices among elements. In this work, we constrain $\gamma$ using observations from AMS-02 and DAMPE within the framework of a spatially dependent propagation model. We establish that for elements with $\left \langle {\mathrm{A}}/{\mathrm{Z}} \right \rangle \approx 2$, the spectral index exhibits a positive correlation with both Z and A. This result may be attributed to differences in fragmentation cross-sections caused by variations in ${\mathrm{A}}$ or ${\mathrm{Z}}$. We predict that the observed spectra of nickel and zinc will be consistent with that of iron, while their injection spectra will be slightly softer than those of iron. We anticipate that further measurements from AMS-02, DAMPE, and HERD will help verify these findings, and we strongly encourage the development of theoretical models to systematically explain this phenomenon.

\end{abstract}

\maketitle

\section{\label{sec:level1}Introduction}

The acceleration mechanism is one of the three fundamental questions in cosmic-ray research, serving as a critical link between the origin and propagation of cosmic rays. It determines the maximum energy achievable by cosmic-ray sources and shapes the spectral form of their injection. Shock acceleration is widely accepted as the standard mechanism for cosmic-ray acceleration. Under ideal conditions, the spectral index $\gamma$ of accelerated particles approximately follows: $\gamma ={3\sigma }/({\sigma -1})$\cite{1988REVIEWS, drury1983introduction, dorman2006cosmic}, where $\sigma$ denotes the shock compression ratio. This implies that different elements should exhibit the same $\gamma$ value at a given rigidity.

However, recent observations from the AMS-02 experiment\cite{aguilar2021alpha,aguilar2024properties,aguilar2023properties,aguilar2021properties,aguilar2021properties1,aguilar2020properties,aguilar2019properties} have revealed systematic differences in the energy spectra of various elements, with particularly notable discrepancies between groups of primary elements—specifically the He-C-O-Fe group and the Ne-Mg-Si-S group\cite{aguilar2021properties}. Dong et al\cite{dong2024new}. suggested that these differences might arise from a higher fraction of secondary contributions in the Ne-Mg-Si-S group. Nevertheless, their study still reported non-negligible variations in $\gamma$ across elements. Similarly, Pan et al\cite{pan2023injection}., in their analysis of AMS-02 spectral data, also concluded that the values of $\gamma$ differ among elements.

A clear tension exists between the theoretical expectation of a universal $\gamma$ and the experimentally observed dependence on elemental species. Although the differences in $\gamma$ could be attributed to variations in elemental origins\cite{pan2023injection}, such an explanation appears contrived and lacks naturalness. Therefore, we aim to systematically investigate the physical principles underlying the divergence in $\gamma$, which may contribute to the advancement of acceleration theory.

In this work, employing a spatially-dependent propagation model (SDP model)\cite{tomassetti2012origin,gaggero2015gamma,guo2015spatial,liu2018revisiting,abeysekara2017extended} and using experimental data from AMS-02 and DAMPE\cite{alemanno2025charge}, we constrain the values of $\gamma$ and find that $\gamma$ increases with atomic mass ${\mathrm{A}}$ or atomic number ${\mathrm{Z}}$. We propose that this trend may be related to the larger fragmentation cross-sections of heavier nuclei, which undergo more significant fragmentation during acceleration, leading to a predictable modification of the spectral index.

\section{MODEL AND METHODOLOGY}

In the following, a quantitative description of the CR propagation process is provided. The propagation of CRs can be represented mathematically by the following partial differential equation\cite{maurin2002galactic}
\begin{equation}
\begin{split}
\frac{\partial\Psi(\vec{r},p,t)}{\partial t} = Q(\vec{r},p,t)+\vec{\nabla}\cdot(D_{xx}\vec{\nabla}\Psi-\vec{V}_{c}\Psi)+ \\
\frac{\partial}{\partial p}[p^{2}D_{pp}\frac{\partial}{\partial p}\frac{\Psi}{p^{2}}]-\frac{\partial}{\partial p}[\dot{p}\Psi-\frac{p}{3}(\vec{\nabla}\cdot\vec{V}_{c})]-\frac{\psi}{\tau_{f}}-\frac{\psi}{\tau_{r}},
\end{split}
\end{equation}

where $Q(\vec{r},p,t)$ describes the distribution of sources,$\vec{V}_{c}$ represents the convection velocity, $D_{pp}$ is the momentum diffusion coefficient accounting for the reacceleration process, $\dot{p}$,$\tau_{f}$ and $\tau_{r}$ are the energy loss rate, the fragmentation timescale, and the radioactive decay timescale, respectively.  
\begin{table}[]
\caption{\label{tab:table1}Propagation model parameters.}
\begin{ruledtabular}
\begin{tabular}{ccccccc}
$D_0{}^{\mathrm{}}[\mathrm{cm}^{-2}\mathrm{s}^{-1}]$ &$\delta_0$&$\quad N_m$&$\quad\xi$&$\quad n$&$\quad v_A[\mathrm{km~s}^{-1}]$&$\quad z_0[\mathrm{kpc}]$\\ \hline
 $5.7\times10^{28}$&0.58\footnotemark[1]&0.24&0.082&4.0&6&5 \\
 %DAMPE&$5.7\times10^{28}$&0.65&0.24&0.082&4.0&5 \\
 
\end{tabular}
\end{ruledtabular}
\footnotetext[1]{When fitting the DAMPE data $\delta_0=0.65$.}
\end{table}
%=6&
\begin{table}[]
\caption{\label{tab:table2}Spectral injection parameters.}
\begin{ruledtabular}

\begin{tabular}{cccccc}
 %\textrm{$\quad\nu_{2}$\footnote{Note a.}}
 &\text{Normalization}\\Element&$[\mathrm{GeV}^{-1}\mathrm{m}^{-2}\mathrm{s}^{-1}\mathrm{sr}^{-1}]$&$\quad\gamma_2$\footnotemark[1]& Abundance \\ \hline
He&$2.54\times 10^{-3}$&$2.34^{+0.01}_{-0.01}$\footnotemark[2]&74550 \\
C&$9.41\times 10^{-5}$&$2.36^{+0.01}_{-0.01}$\footnotemark[3]&2760 \\
N&$5.63\times 10^{-6}$&$2.34^{+0.01}_{-0.01}$&165 \\
O&$1.23\times 10^{-4}$&$2.37^{+0.02}_{-0.02}$\footnotemark[4]&3600 \\
Ne&$1.79\times 10^{-5}$&$2.38^{+0.02}_{-0.02}$&526 \\
Na&$5.45\times 10^{-7}$&$2.40^{+0.01}_{-0.01}$&16 \\
Mg&$2.33\times 10^{-5}$&$2.40^{+0.02}_{-0.02}$&684 \\
Al&$1.67\times 10^{-6}$&$2.39^{+0.01}_{-0.01}$&49 \\
Si&$2.36\times 10^{-5}$&$2.40^{+0.02}_{-0.02}$&691 \\
S&$3.89\times 10^{-6}$&$2.41^{+0.02}_{-0.02}$&114 \\
Fe&$2.50\times 10^{-5}$&$2.42^{+0.03}_{-0.02}$\footnotemark[5]&734 \\

\end{tabular}
\end{ruledtabular}
\footnotetext[1]{$\gamma_2$ is constrained by $\chi^2/\mathrm{d.o.f.} < 2$, $\gamma_1=2.10$, $R_{br}=8$GV.}
\footnotetext[2]{When fitting the DAMPE data $\gamma_2=2.32^{+0.01}_{-0.01}$.}
\footnotetext[3]{When fitting the DAMPE data $\gamma_2=2.33^{+0.01}_{-0.01}$.}
\footnotetext[4]{When fitting the DAMPE data $\gamma_2=2.35^{+0.01}_{-0.01}$.}
\footnotetext[5]{When fitting the DAMPE data $\gamma_2=2.40^{+0.01}_{-0.02}$.}
\end{table}

The SDP model divides the cosmic ray diffusion region into two parts: the inner zone is a cylindrical area centered on the galactic disk plane and extending a certain thickness on both sides, while the remaining areas constitute the outer zone. The diffusion speed of CRs in the inner zone is lower than that in the outer zone and depends on the distribution of sources. The diffusion coefficient of the model is expressed as:
\begin{figure*}[htbp]  % 添加浮动位置参数
    \centering
    % 第一行子图
    \begin{subfigure}[b]{0.48\textwidth}  % [b]底部对齐
        \centering
        \includegraphics[
            width=\linewidth,
            height=7cm,                  % 固定高度
            keepaspectratio,             % 保持比例
            bb=0 0 500 400,              % 手动设置BoundingBox (单位: bp)
            %clip                         % 裁剪溢出部分
        ]{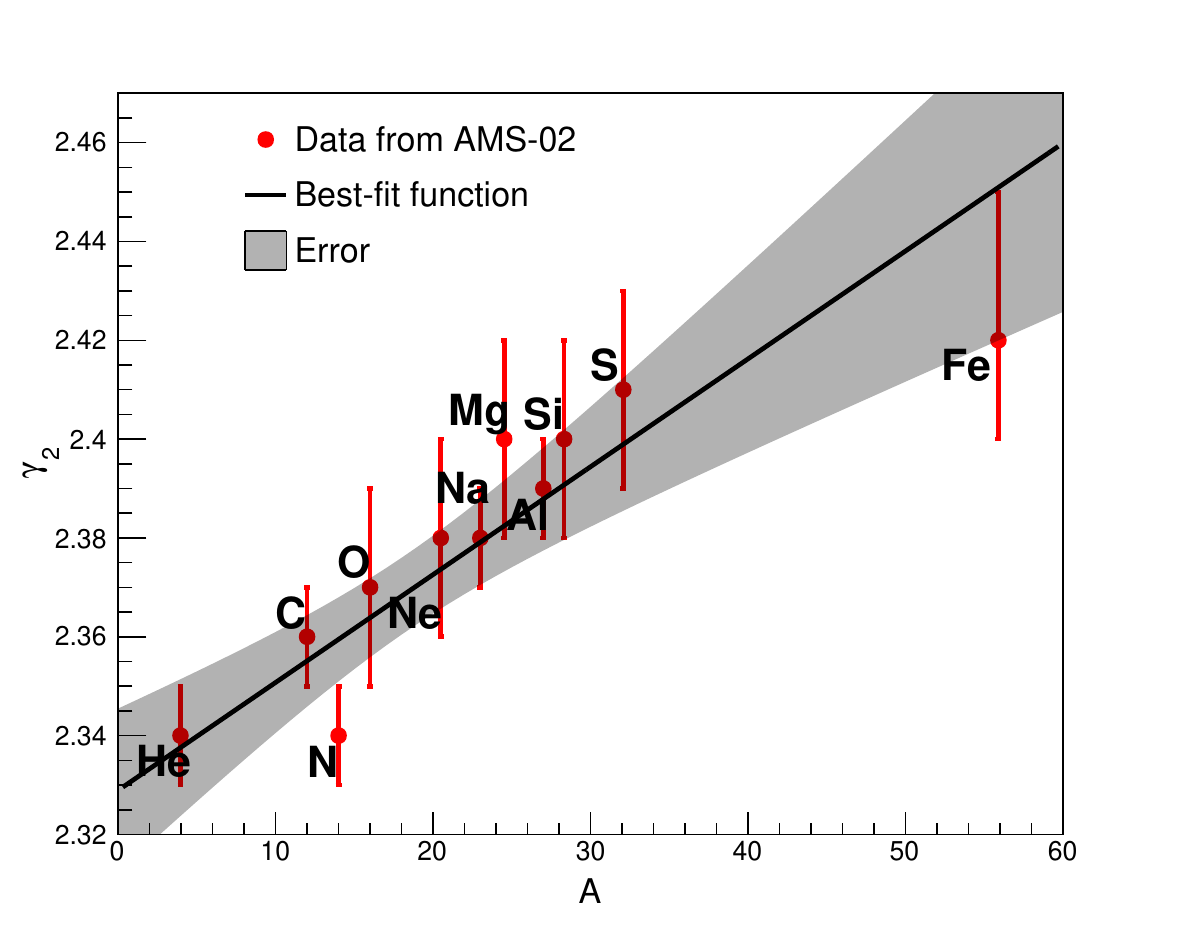}
      %  \caption{$\gamma_2$ vs A (AMS-02)}
    \end{subfigure}
    \hfill
    \begin{subfigure}[b]{0.48\textwidth}
        \centering
        \includegraphics[
            width=\linewidth,
            height=7cm,
            bb=0 0 500 400,              % 统一所有图片的BB值
           % clip
        ]{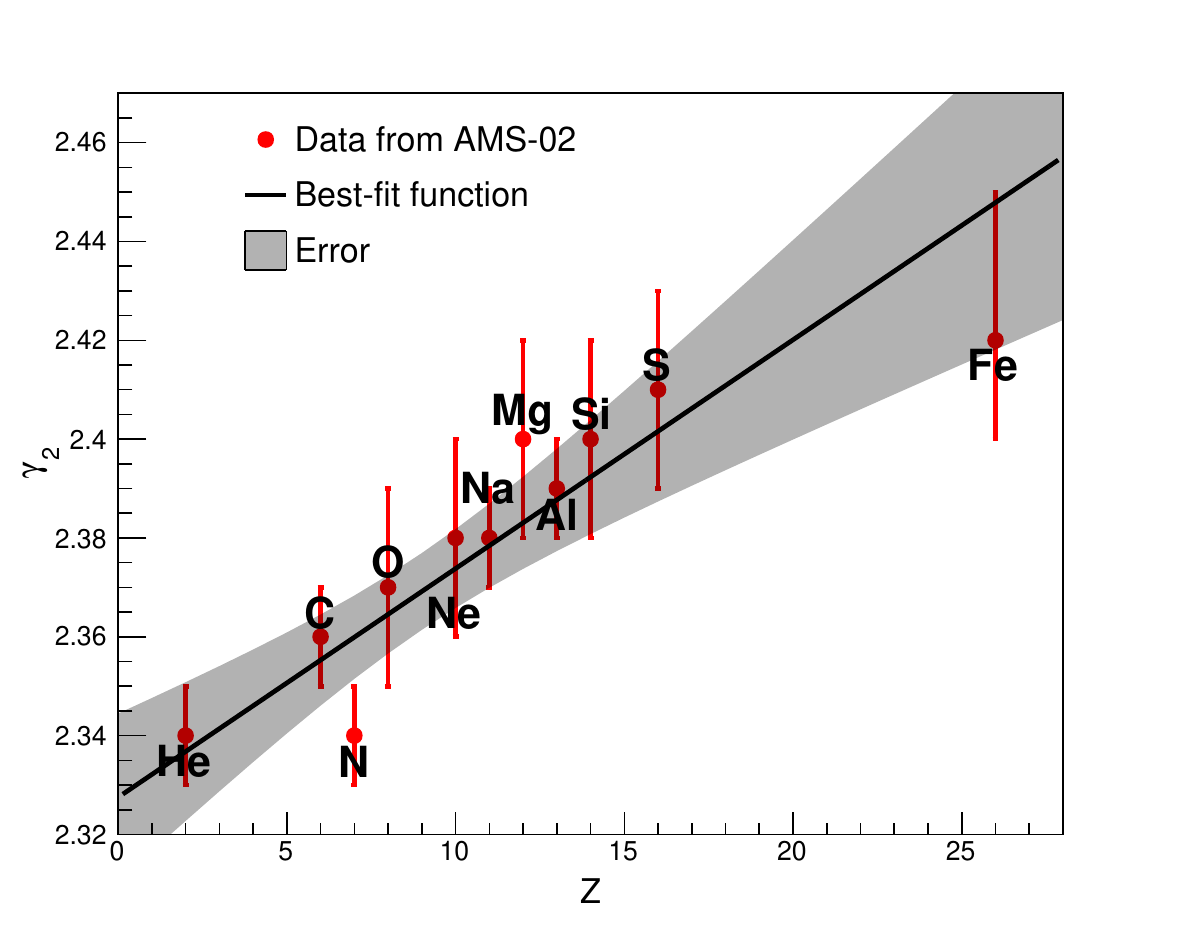}
      %  \caption{$\gamma_2$ vs Z (AMS-02)}
    \end{subfigure}
    
    % 第二行子图（垂直间距调整）
    \vspace{0.6cm}  % 增加行间距
    
    \begin{subfigure}[b]{0.48\textwidth}
        \centering
        \includegraphics[
            width=\linewidth,
            height=7cm,
            bb=0 0 500 400,
          %  clip
        ]{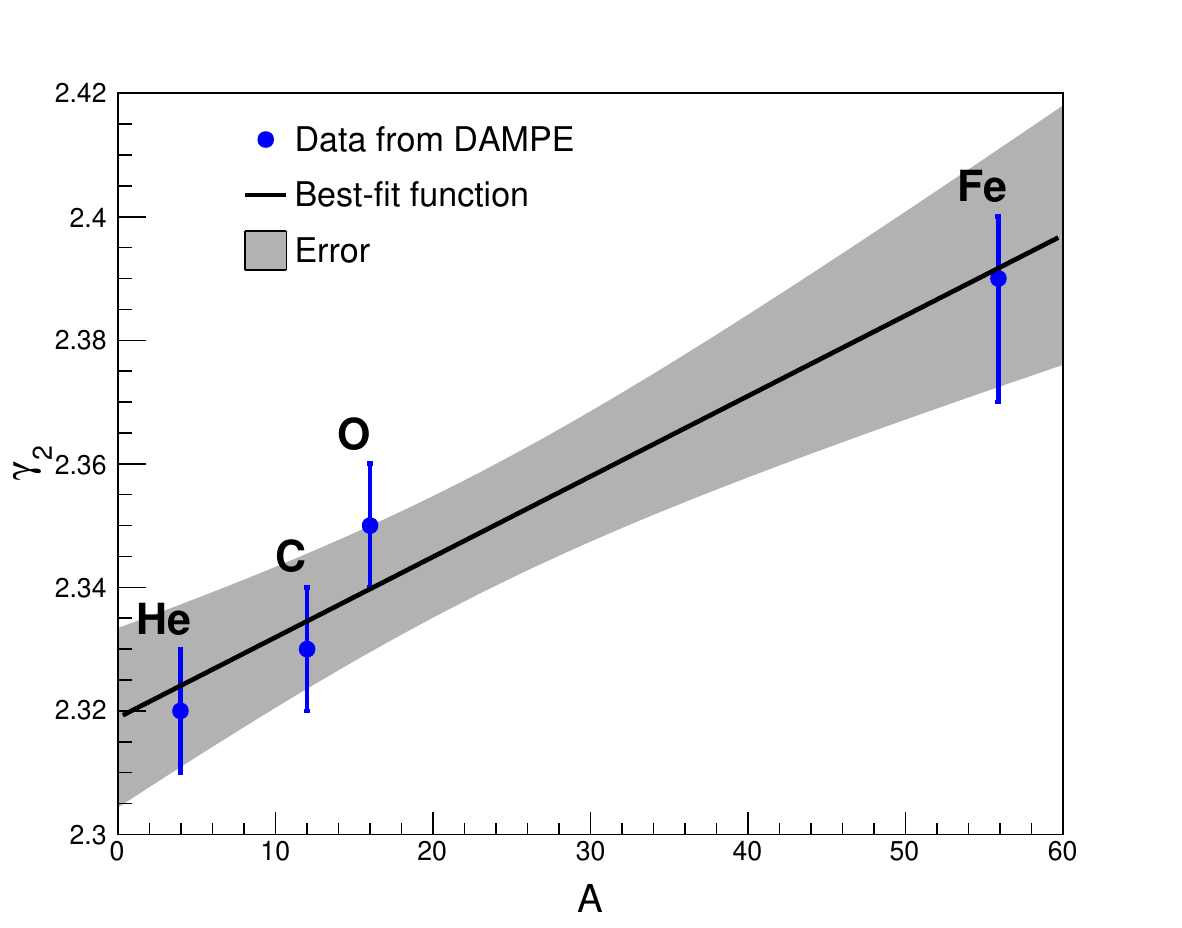}
    %    \caption{$\gamma_2$ vs A (DAMPE)}
    \end{subfigure}
    \hfill
    \begin{subfigure}[b]{0.48\textwidth}
        \centering
        \includegraphics[
            width=\linewidth,
            height=7cm,
            bb=0 0 500 400,
        %    clip
        ]{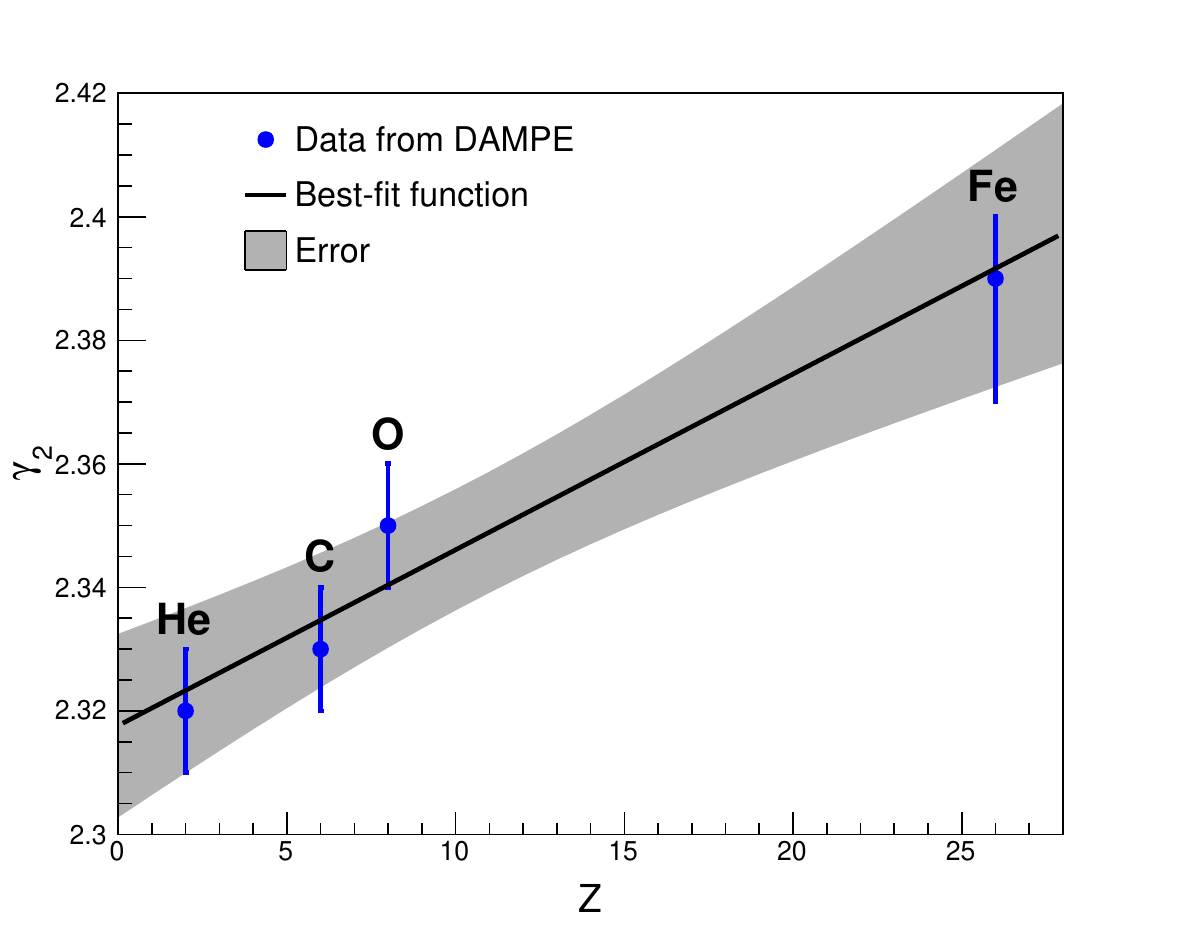}
    %    \caption{$\gamma_2$ vs Z (DAMPE)}
    \end{subfigure}
    
    \caption{Correlation between the injection spectral index $\gamma_2$ and the nucleon number ${\mathrm{A}}$ or proton number ${\mathrm{Z}}$ for various elements. The red and blue data points represent the fitting results derived from the AMS-02 and DAMPE data, respectively. The uncertainties in $\gamma_2$ were determined by constraining $\chi^2/\mathrm{d.o.f.} < 2$. The solid red and blue lines correspond to the linear fits to the AMS-02 and DAMPE data, respectively, with the shaded areas indicating the associated uncertainties of the fits.
Top left: $\gamma_2$ as a function of ${\mathrm{A}}$ from AMS-02;
Top right: $\gamma_2$ as a function of ${\mathrm{Z}}$ from AMS-02;
Bottom left: $\gamma_2$ as a function of ${\mathrm{A}}$ from DAMPE;
Bottom right: $\gamma_2$ as a function of ${\mathrm{Z}}$ from DAMPE.   }
    \label{fig:spectral_correlation}  % 修改为更具体的标签
\end{figure*}

\begin{equation}D_{xx}(r,z,R)=D_{0}F(r,z)\beta^{\eta}(\frac{R}{R_{0}})^{\delta(r,z)},\end{equation}
where the function $F(r,z)$ is defined as:
\begin{equation}F(r,z)=\begin{cases} g(r,z)+[1-g(r,z)]\left(\frac{z}{\xi z_0}\right)^n,&|z|\le\xi z_0\\ 1,&|z|>\xi z_0\end{cases},\end{equation}
With $g(r,z)=N_m/[1+f(r,z)]$. Here the diffusion coefficient of the inner zone is anticorrelated with the source distribution $f(r,z)$ , given by
\begin{equation}f(r,z)=\left(\frac{r}{r_\odot}\right)^{1.25}\exp\biggl[-\frac{3.87(r-r_\odot)}{r_\odot}\biggr]\exp\biggl(-\frac{|z|}{z_s}\biggr),\end{equation} 
where $r_{\odot}=8.5\text{ kpc}$ and $z_s=0.2\text{ kpc}$ . For the outer zone, the diffusion coefficient remains constant when varying spatial locations.In contrast, the diffusion coefficient in the traditional model remains constant throughout the entire space. 

The source spectrum of CRs is assumed to be a broken power law in rigidity. 
\begin{equation}q\left ( R \right ) \propto\begin{cases}&\left (\frac{R}{R_{br}}\right ) ^{-\gamma _{1} }  ,R< R_{br} \\ &\left (\frac{R}{R_{br}}  \right ) ^{-\gamma _{2} } ,R\ge R_{br}\end{cases}\end{equation}
To solve the transport equation, we employ the numerical package DRAGON\cite{evoli2008cosmic} . The force field approximation is incorporated to account for solar modulation effects\cite{gleeson1968solar}. The key parameters related to the SDP model are summarized in Table I, while the spectral parameters of each element are summarized in Table II. Since there are slight differences in the spectral indices measured by the AMS-02 and DAMPE experiments, we performed separate fits to the data from both experiments, adjusted some parameters accordingly, and marked the corresponding entries in the tables.

\section{RESULTS}
Based on the above configuration, our calculations successfully reproduce the energy spectral fluxes of He, C, O, Ne, Mg, Si, S, Fe, N, Na, and Al as measured by AMS-02, as well as those of He, C, O, and Fe obtained by DAMPE. To minimize potential influences from solar modulation effects, we focus our analysis on the relationship between $\gamma _{2}$ and the atomic mass number ${\mathrm{A}}$ as well as the atomic number ${\mathrm{Z}}$. The specific results are shown in Figure 1.

The top-left panel of Figure 1 displays the relationship between $\gamma _{2}$ and ${\mathrm{A}}$ derived from the AMS-02 data, with a correlation coefficient of $r = 0.73^{+0.17}_{-0.17}$ ($p = \SI{0.005}{}$). The linear fit yields: 
\begin{equation}\gamma_{2} = 0.0022_{+0.0004}^{-0.0004}  {\mathrm{A}} +2.329_{-0.008}^{+0.008} \end{equation}, with $\chi^2/\mathrm{d.o.f.} = 6.70/9$. The top-right panel shows the relationship between $\gamma _{2}$ and ${\mathrm{Z}}$ based on the AMS-02 data, also exhibiting a correlation coefficient of $r = 0.73^{+0.17}_{-0.17}$ ($p = \SI{0.0043}{}$). The linear regression result is: 
\begin{equation}\gamma_{2} = 0.0046_{+0.0009}^{-0.0009} {\mathrm{Z}} +2.328_{-0.009}^{+0.009}\end{equation}, with $\chi^2/\mathrm{d.o.f.} = 6.45/9$. The bottom-left panel presents the relationship between $\gamma _{2}$ and ${\mathrm{A}}$ from the DAMPE dataset, which similarly gives a correlation coefficient of $r = 0.72^{+0.18}_{-0.18}$ ($p = \SI{0.582}{}$). The linear fit result is: 
\begin{equation}\gamma_{2} =0.0013_{+0.0002}^{-0.0002}  {\mathrm{A}} -2.319_{-0.007}^{+0.007}  \end{equation}
, with $\chi^2/\mathrm{d.o.f.} = 1.45/2$. The bottom-right panel illustrates the relationship between $\gamma _{2}$ and ${\mathrm{Z}}$ based on the DAMPE data, with a correlation coefficient of $r = 0.72^{+0.18}_{-0.18}$ ($p = \SI{0.575}{}$). The linear regression gives: 
\begin{equation}\gamma_{2} = 0.0028_{+0.0005}^{-0.0005} {\mathrm{Z}} +2.318_{-0.008}^{+0.008}\end{equation}, with $\chi^2/\mathrm{d.o.f.} = 1.28/2$.

Taken together, these results clearly indicate a significant positive correlation between $\gamma _{2}$ and both ${\mathrm{A}}$ and ${\mathrm{Z}}$ for elements with $ {\mathrm{A}}/{\mathrm{Z}} \approx 2$. We speculate that this phenomenon may be related to the fragmentation effect during the acceleration process: nuclei with larger mass numbers ${\mathrm{A}}$ possess larger fragmentation cross-sections, leading to greater losses due to fragmentation during acceleration, which ultimately softens the energy spectrum. Alternatively, this result could also stem from differences in energy loss during the injection and acceleration processes among different elements.

\section{DISCUSSION}

The results of this study cannot yet account for the softer proton injection spectrum. This phenomenon is generally attributed to the injection rate of particles into the acceleration region\cite{hanusch2017anomalies} . Based on the results presented here, we anticipate $\gamma_{\mathrm{Ni}}\approx\gamma_{\mathrm{Zn}}\approx2.44$\cite{adriani2024direct}. Given that ${\mathrm{Ni}}$ and ${\mathrm{Zn}}$ contain no secondary components, their energy spectra are expected to be similar to that of iron. The currently limited number of elemental energy spectra measured by DAMPE leads to an elevated p-value. Therefore, we anticipate that future observations from AMS-02, DAMPE, and HERD will provide more data to further verify and confirm this result.

\section{Conclusions}
Based on observational data from AMS-02 and DAMPE, this study confirms a positive correlation between the injection spectral index $\gamma$ of various elements in cosmic rays and both the nucleon number ${\mathrm{A}}$ and the proton number ${\mathrm{Z}}$. We hypothesize that this relationship may be associated with collisional fragmentation effects experienced by elements during the acceleration process.

\begin{acknowledgments}
We gratefully acknowledge Lin Nie, Zigui Dong, and others for their substantial assistance and valuable suggestions in the completion of this work. This work was financially supported by the National Key R\&D Program of China (Grant No. 2025SKA0110103), the National Natural Science Foundation of China (Grants No. 12333006, 12275279, 12373105, and 12320101005), and the  Postgraduate Innovation Fund Project of Hebei Provincial Education Department (Grant No. CXZZBS2026083).

%This work is supported in China by National Key R&D Program of
%China under the grant 2025SKA0110103 the National Natural Science
%Foundation of China (Nos.12333006, 12275279, 12373105, 12320101005) and Postgraduate Innovation Fund Project of Hebei Provincial Education Department (No. CXZZBS2026083).
\end{acknowledgments}
\appendix
\section{The Injection Spectra and Energy Spectra in the SDP Model}

\begin{figure*}[htbp]  % 添加浮动位置参数
    \centering
    % 左子图（AMS-02数据）
    \begin{subfigure}[t]{0.48\textwidth}  % [t]顶部对齐
        \centering
        \includegraphics[
            width=\linewidth,
            height=7cm,                  % 固定高度（根据内容调整）
            keepaspectratio,             % 保持原始比例
            bb=0 0 500 400,              % A4尺寸的BoundingBox（单位：bp）
       %     clip,                        % 裁剪溢出部分
            draft=false                  % 确保渲染高清图片
        ]{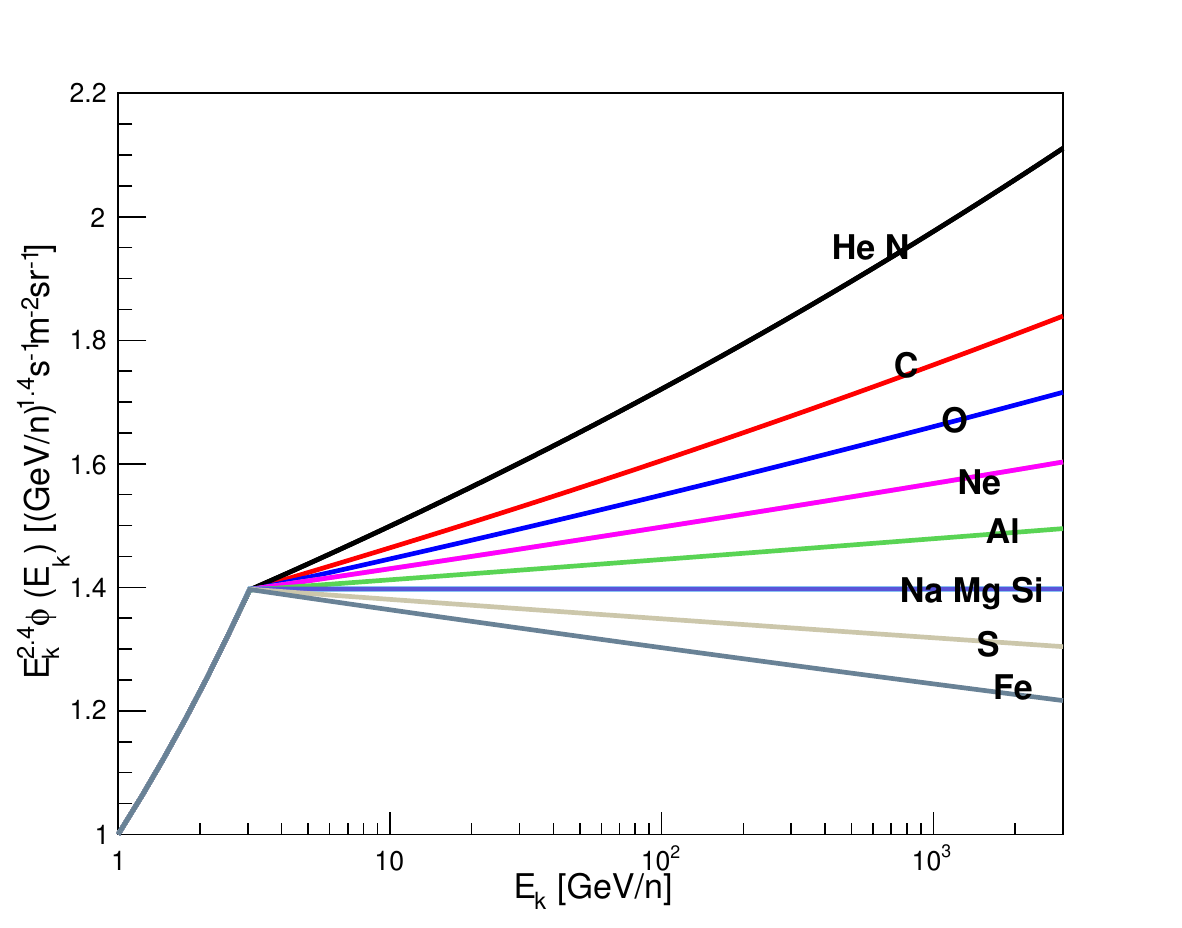}
     %   \caption{Injection spectra (AMS-02)}
        \label{fig:inj_ams}
    \end{subfigure}
    \hfill
    % 右子图（DAMPE数据）
    \begin{subfigure}[t]{0.48\textwidth}
        \centering
        \includegraphics[
            width=\linewidth,
            height=7cm,                  % 与左图等高
            bb=0 0 500 400,              % 统一BoundingBox
        %    clip
        ]{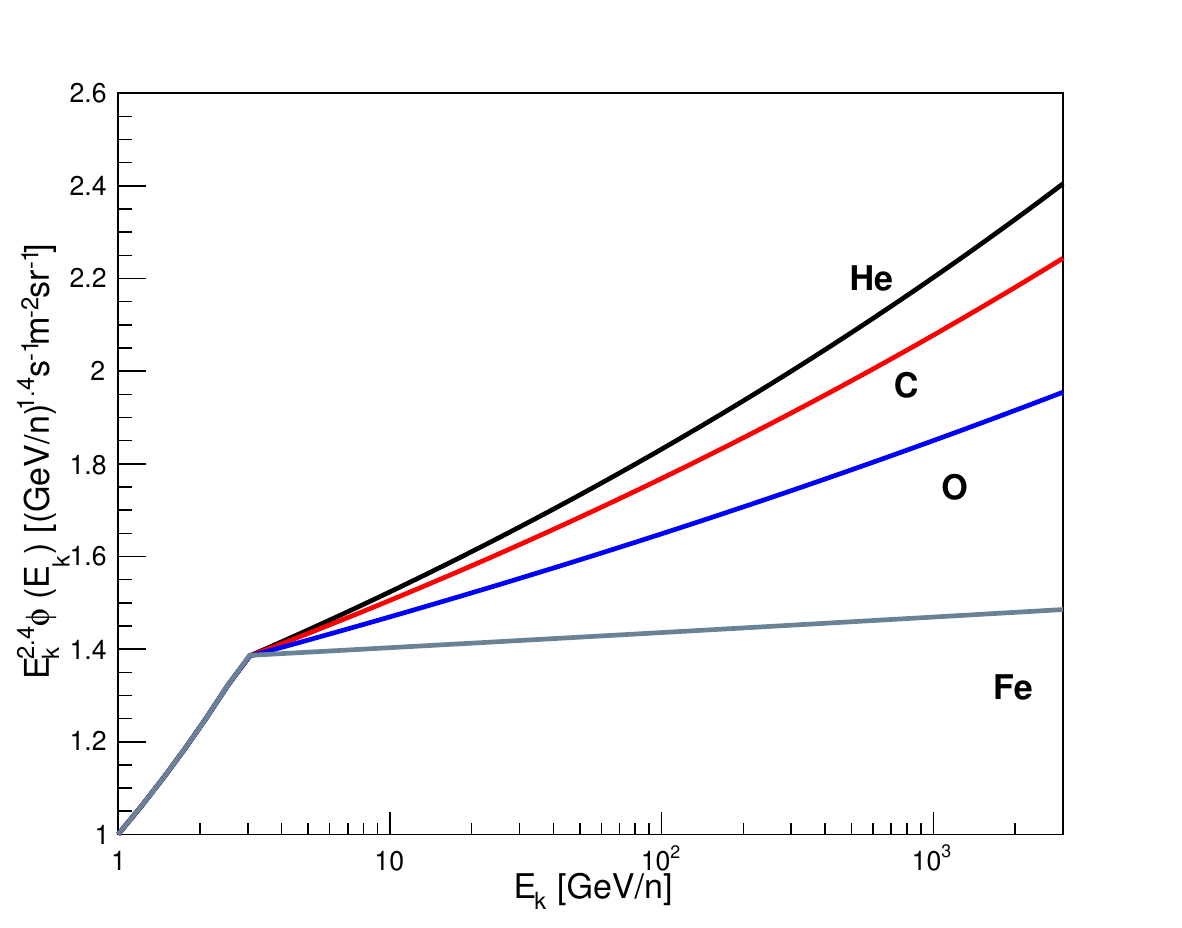}
       % \caption{Injection spectra (DAMPE)}
        \label{fig:inj_dampe}
    \end{subfigure}
    
    \caption{The injection energy spectra of various elements derived from the spatially dependent propagation (SDP) model fitting to AMS-02 data (left) and DAMPE data (right).   }
    \label{fig:injection_spectra}
\end{figure*}
\begin{figure*}[htbp]  % 添加浮动位置参数
    \centering
    % 第一行子图
    \begin{subfigure}[t]{0.24\textwidth}  % [t]顶部对齐
        \centering
        \includegraphics[
            width=\linewidth,
            height=3.5cm,                % 修改为3.5cm固定高度
            keepaspectratio,             % 保持原始比例
            bb=0 0 500 400,              % A4尺寸的BoundingBox（单位：bp）
            draft=false                  % 确保渲染高清图片
        ]{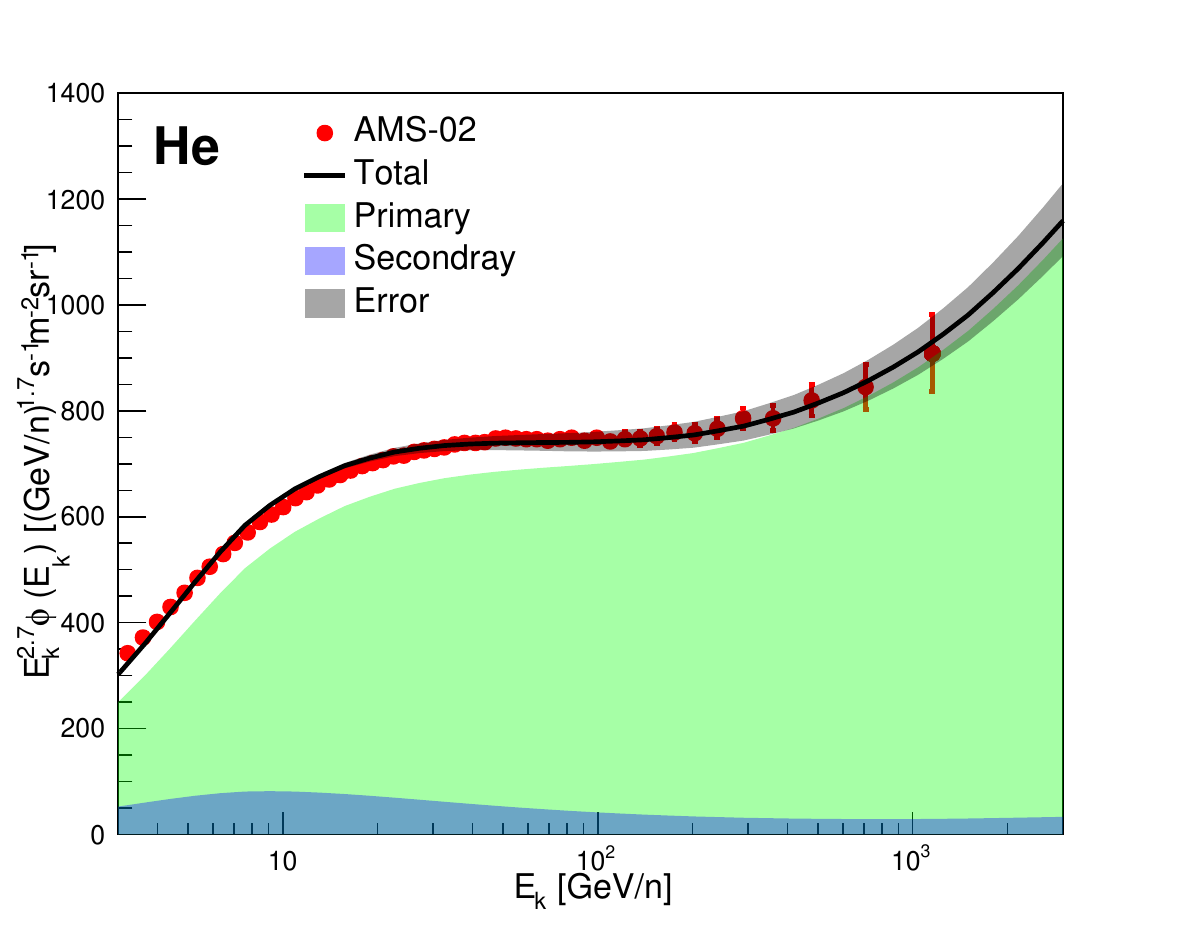}
        \label{fig:he_ams}
    \end{subfigure}
    \hfill
    \begin{subfigure}[t]{0.24\textwidth}
        \centering
        \includegraphics[
            width=\linewidth,
            height=3.5cm,
            bb=0 0 500 400,
            draft=false
        ]{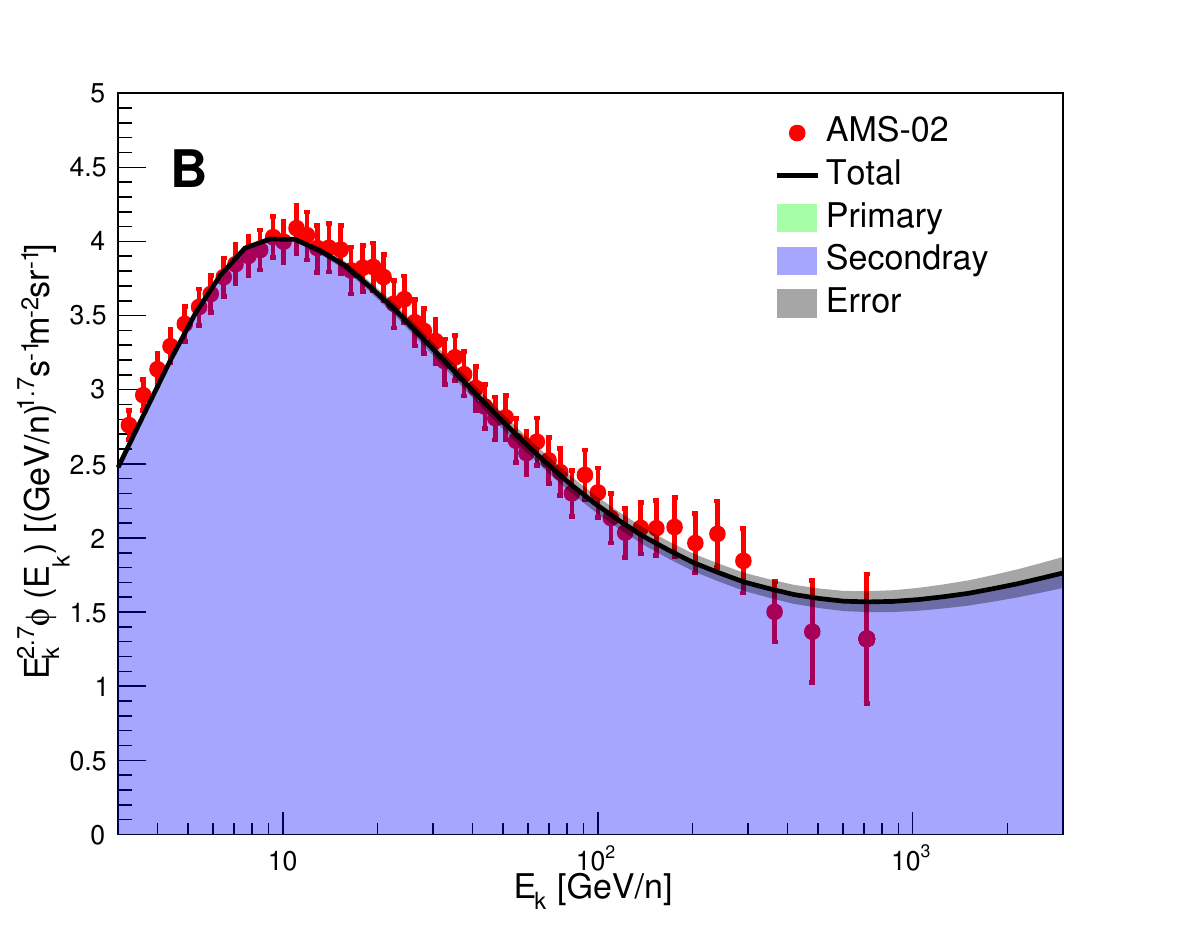}
        \label{fig:b_ams}
    \end{subfigure}
    \hfill
    \begin{subfigure}[t]{0.24\textwidth}
        \centering
        \includegraphics[
            width=\linewidth,
            height=3.5cm,
            bb=0 0 500 400,
            draft=false
        ]{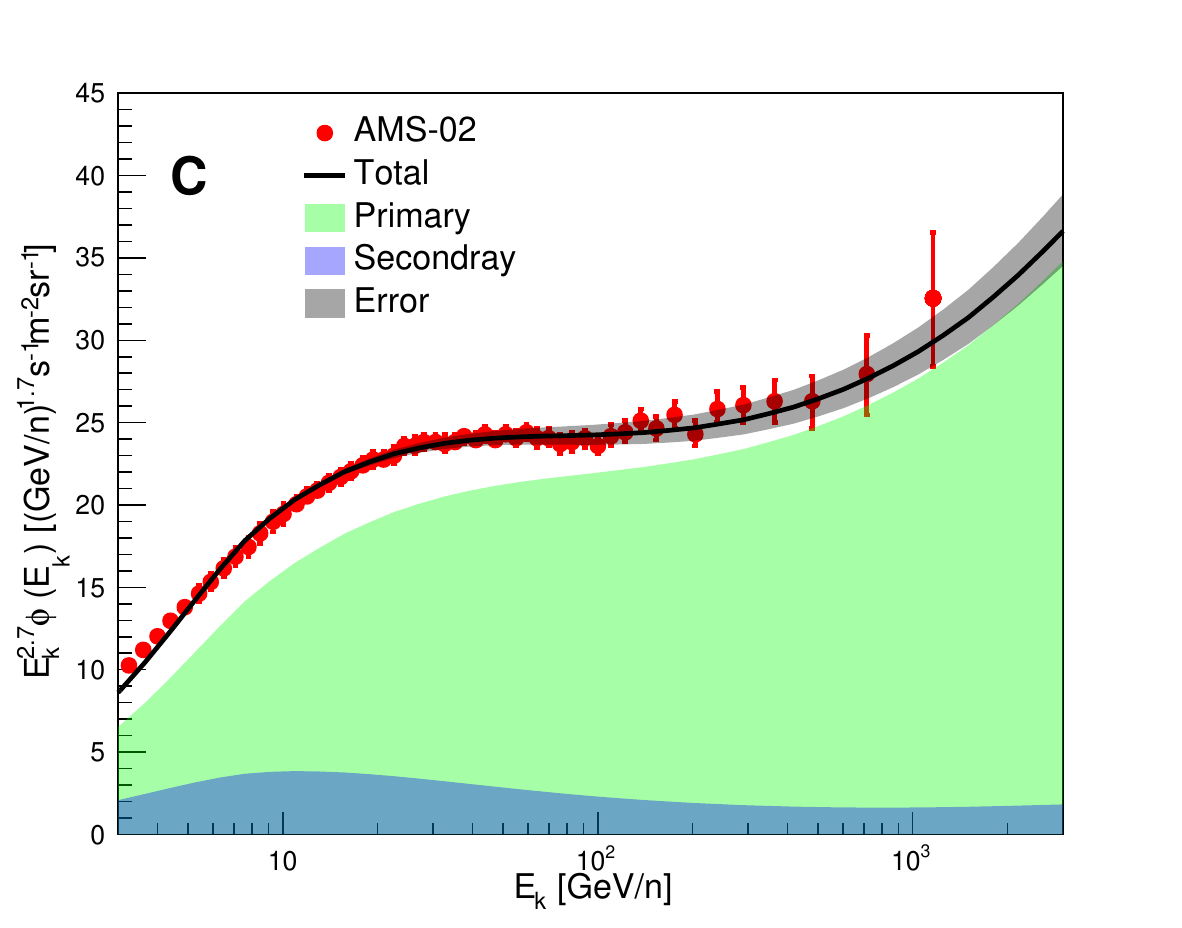}
        \label{fig:c_ams}
    \end{subfigure}
    \hfill
    \begin{subfigure}[t]{0.24\textwidth}
        \centering
        \includegraphics[
            width=\linewidth,
            height=3.5cm,
            bb=0 0 500 400,
            draft=false
        ]{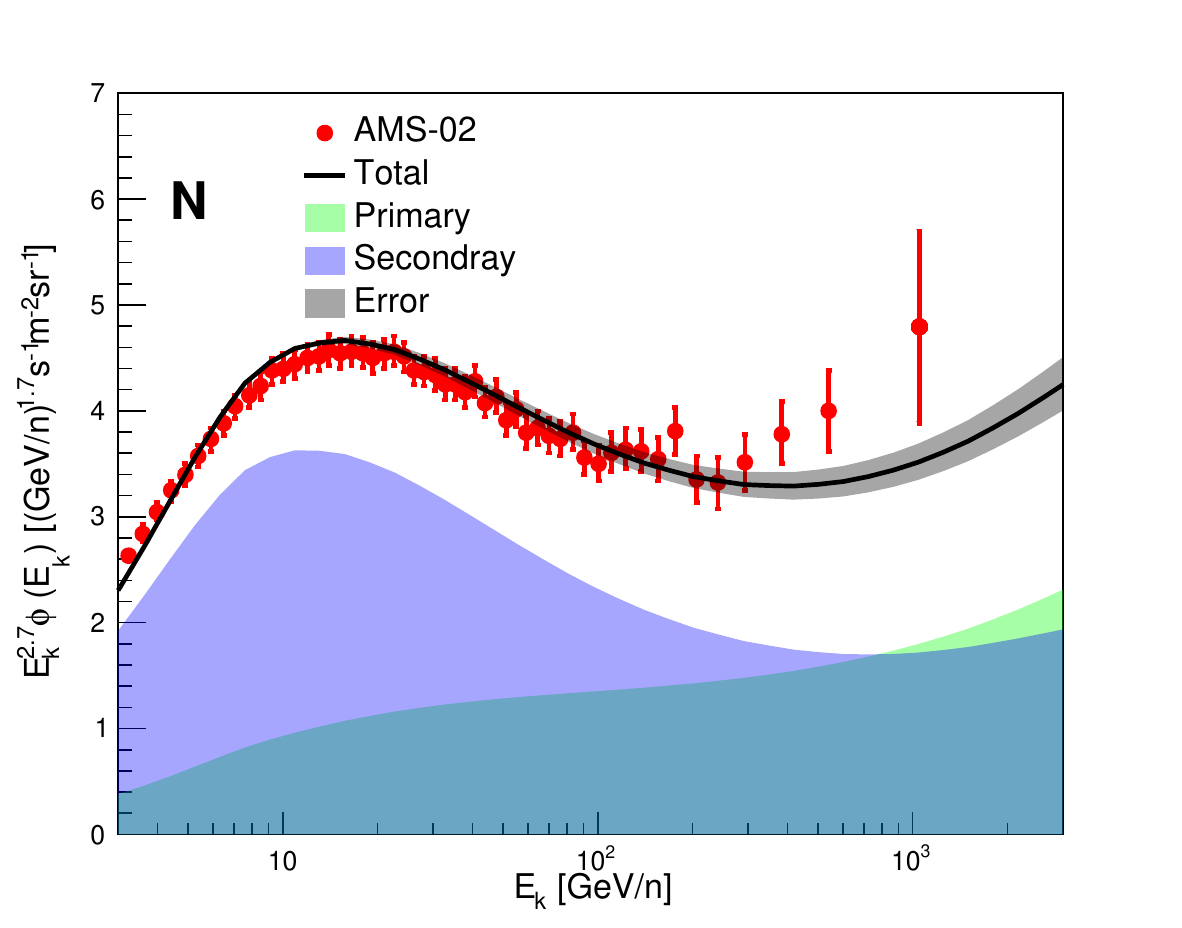}
        \label{fig:n_ams}
    \end{subfigure}
    
    % 第二行子图（添加垂直间距）
    \vspace{0.5cm}
    
    \begin{subfigure}[t]{0.24\textwidth}
        \centering
        \includegraphics[
            width=\linewidth,
            height=3.5cm,
            bb=0 0 500 400,
            draft=false
        ]{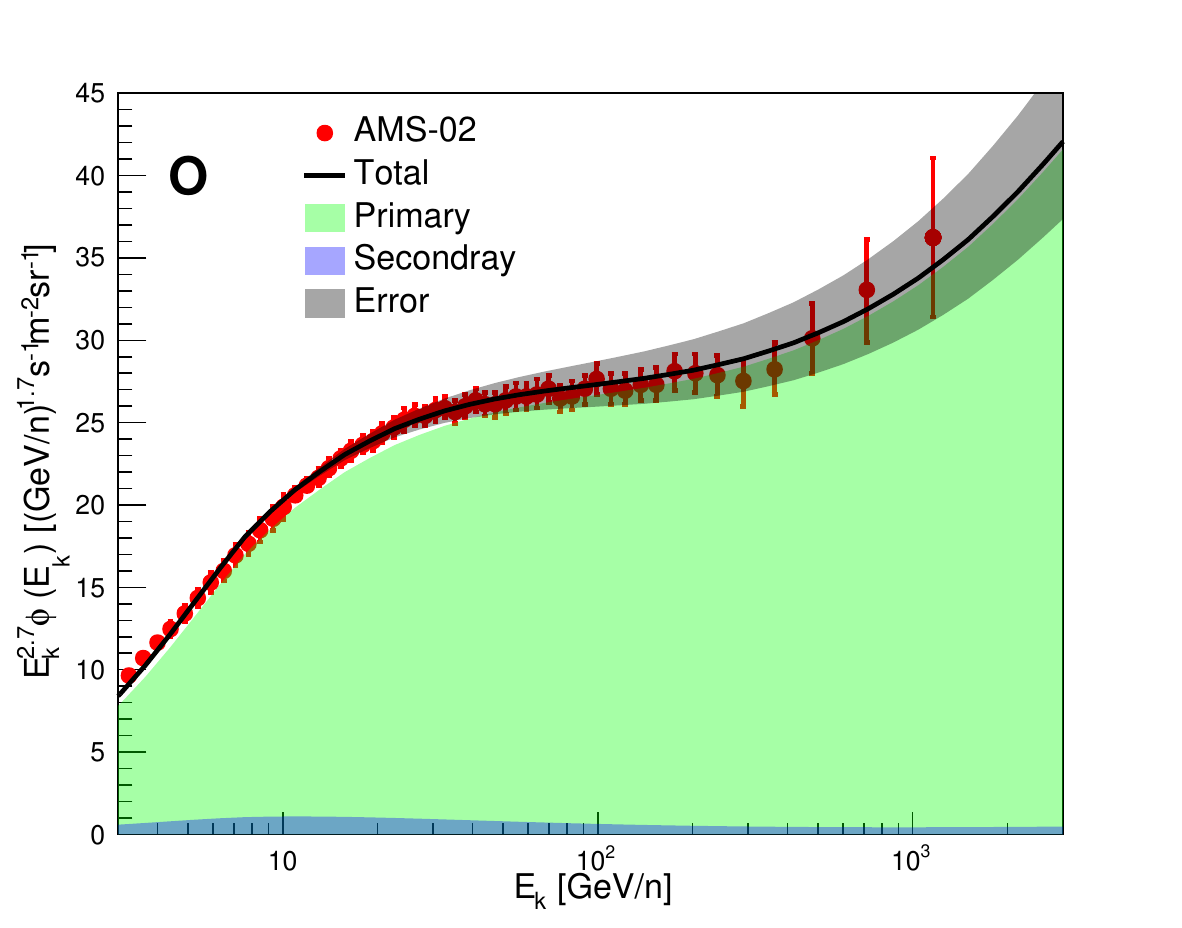}
        \label{fig:o_ams}
    \end{subfigure}
    \hfill
    \begin{subfigure}[t]{0.24\textwidth}
        \centering
        \includegraphics[
            width=\linewidth,
            height=3.5cm,
            bb=0 0 500 400,
            draft=false
        ]{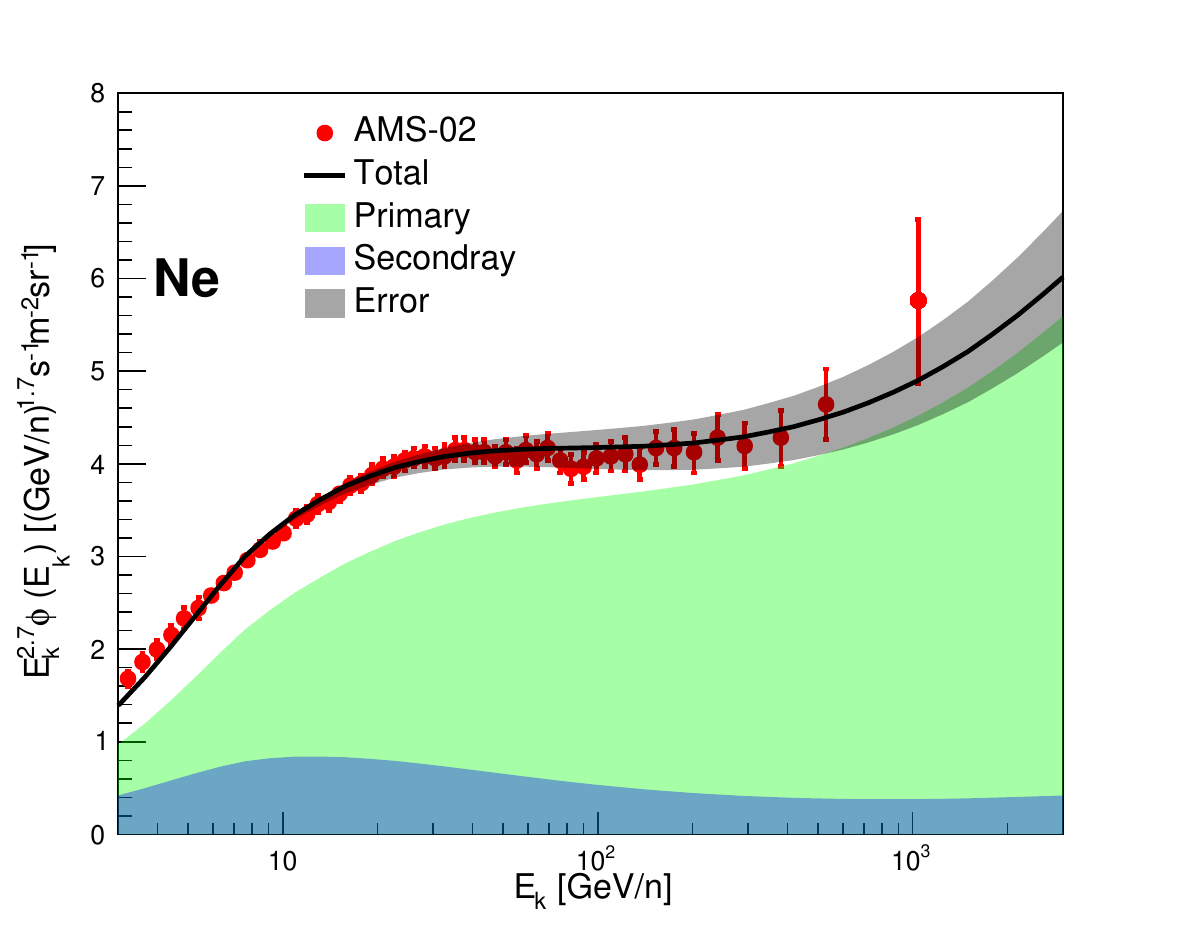}
        \label{fig:ne_ams}
    \end{subfigure}
    \hfill
    \begin{subfigure}[t]{0.24\textwidth}
        \centering
        \includegraphics[
            width=\linewidth,
            height=3.5cm,
            bb=0 0 500 400,
            draft=false
        ]{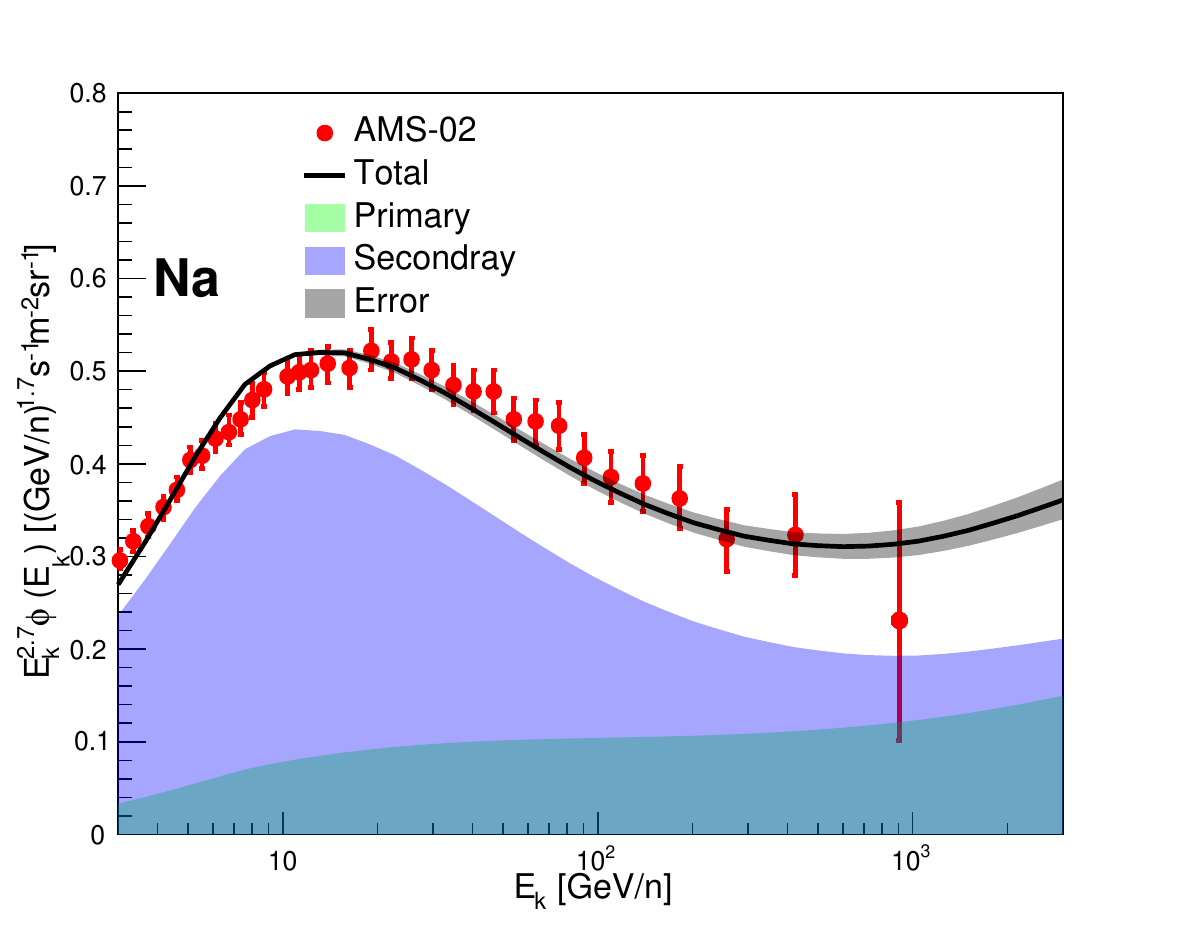}
        \label{fig:na_ams}
    \end{subfigure}
    \hfill
    \begin{subfigure}[t]{0.24\textwidth}
        \centering
        \includegraphics[
            width=\linewidth,
            height=3.5cm,
            bb=0 0 500 400,
            draft=false
        ]{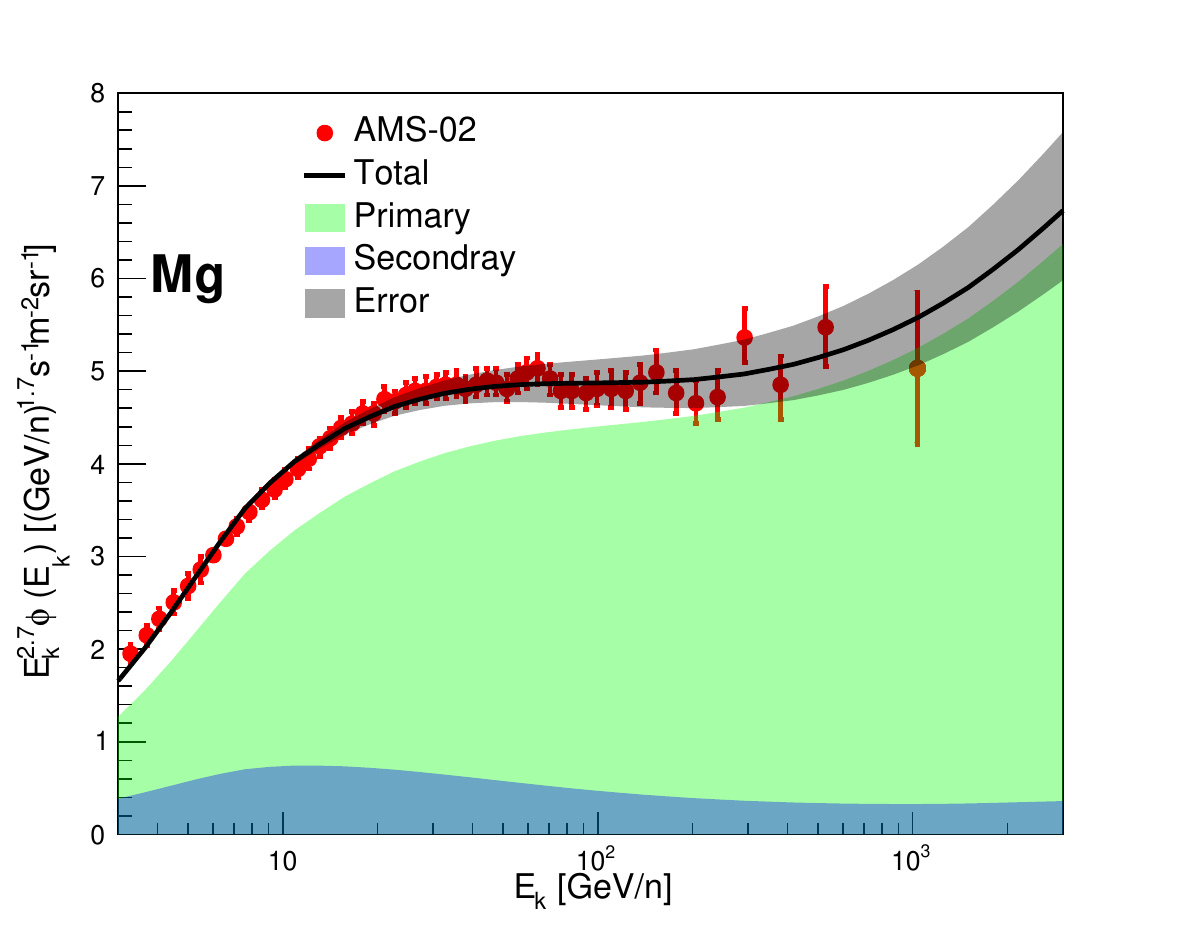}
        \label{fig:mg_ams}
    \end{subfigure}
    
    % 第三行子图
    \vspace{0.5cm}
    
    \begin{subfigure}[t]{0.24\textwidth}
        \centering
        \includegraphics[
            width=\linewidth,
            height=3.5cm,
            bb=0 0 500 400,
            draft=false
        ]{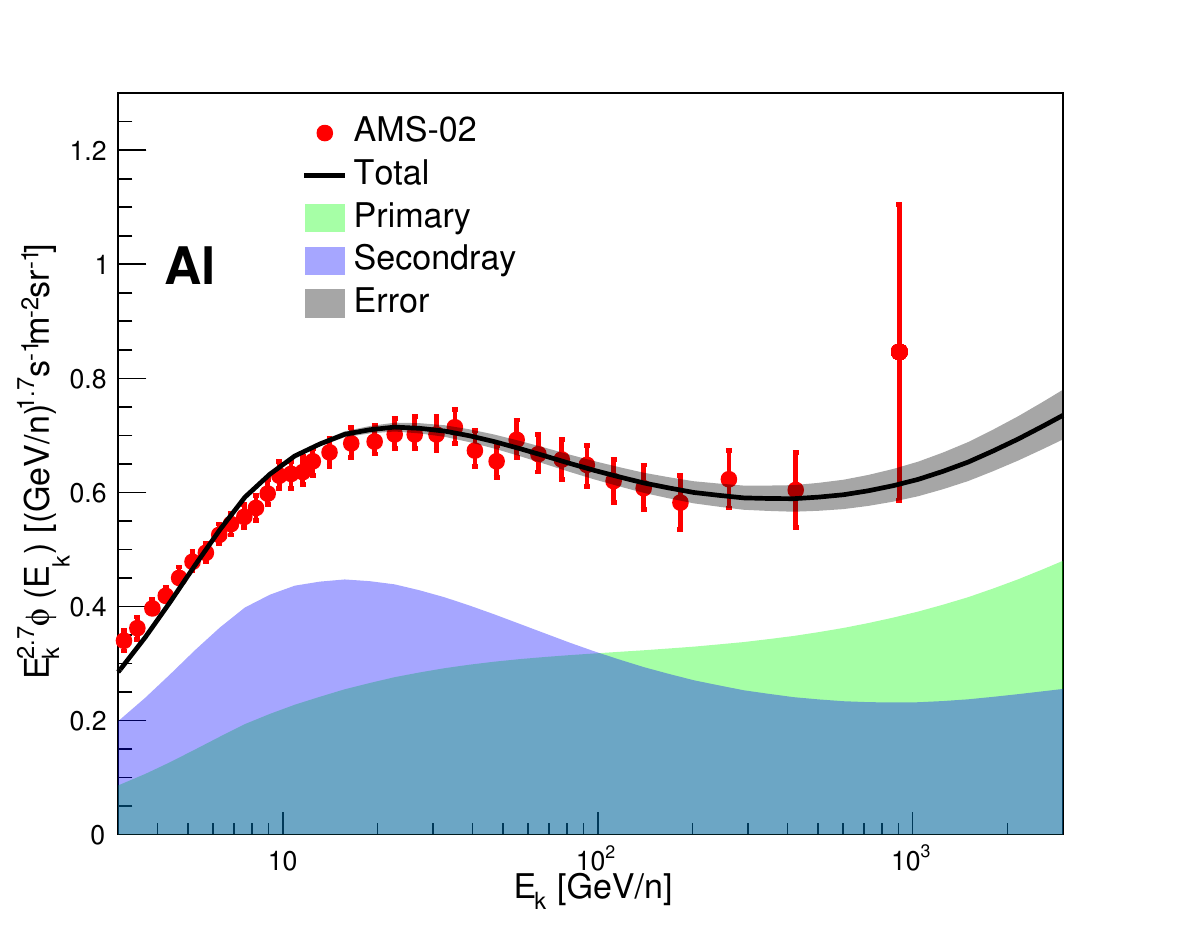}
        \label{fig:al_ams}
    \end{subfigure}
    \hfill
    \begin{subfigure}[t]{0.24\textwidth}
        \centering
        \includegraphics[
            width=\linewidth,
            height=3.5cm,
            bb=0 0 500 400,
            draft=false
        ]{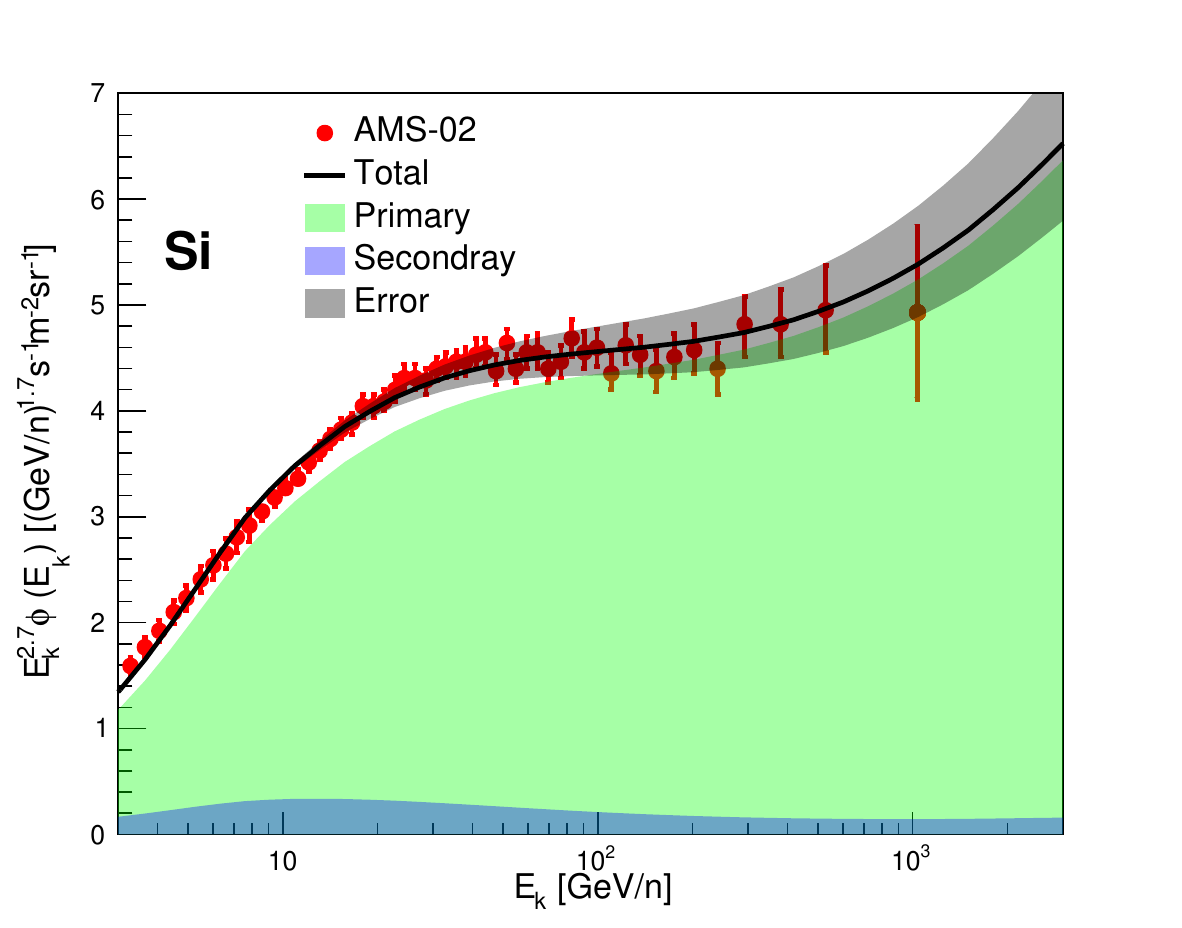}
        \label{fig:si_ams}
    \end{subfigure}
    \hfill
    \begin{subfigure}[t]{0.24\textwidth}
        \centering
        \includegraphics[
            width=\linewidth,
            height=3.5cm,
            bb=0 0 500 400,
            draft=false
        ]{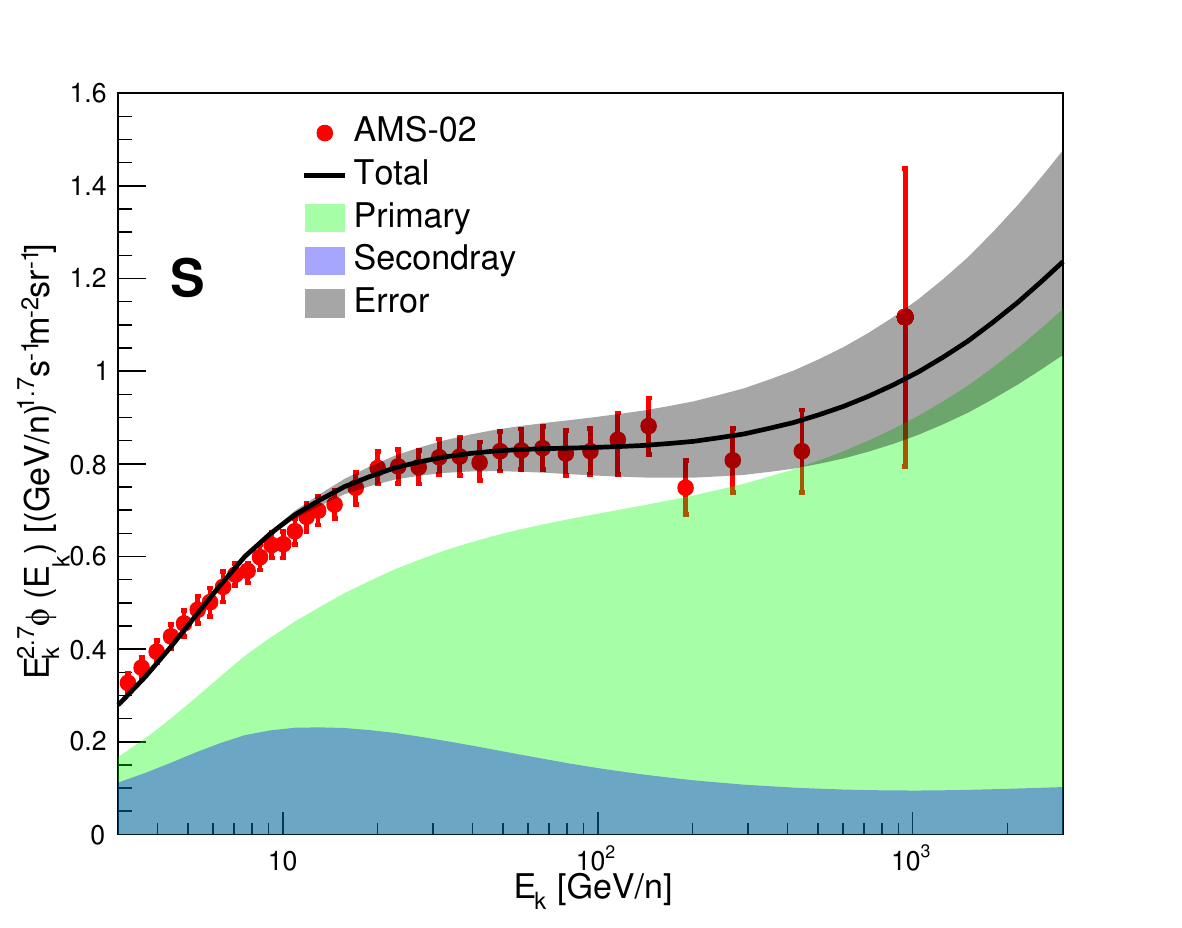}
        \label{fig:s_ams}
    \end{subfigure}
    \hfill
    \begin{subfigure}[t]{0.24\textwidth}
        \centering
        \includegraphics[
            width=\linewidth,
            height=3.5cm,
            bb=0 0 500 400,
            draft=false
        ]{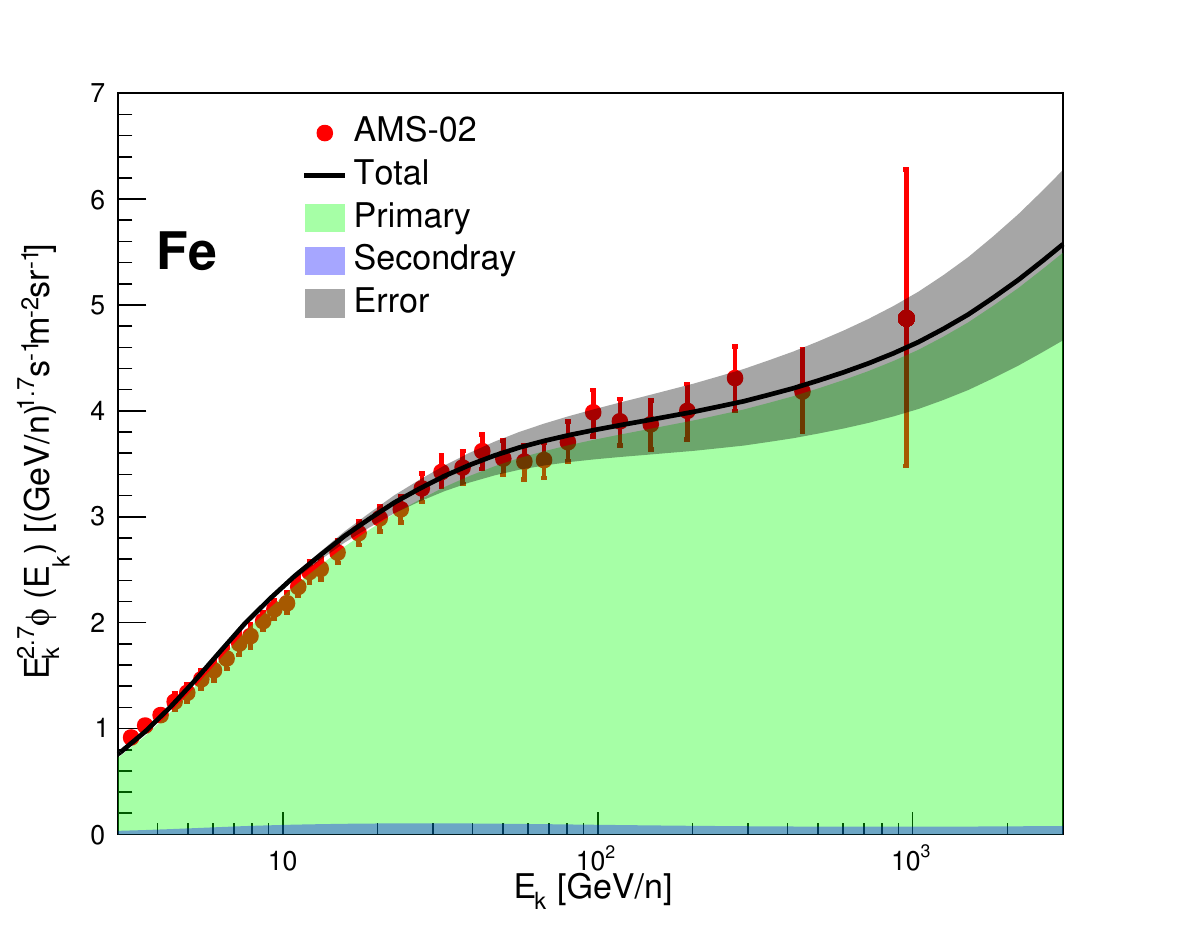}
        \label{fig:fe_ams}
    \end{subfigure}
    
    % 第四行子图
    \vspace{0.5cm}
    
    \begin{subfigure}[t]{0.24\textwidth}
        \centering
        \includegraphics[
            width=\linewidth,
            height=3.5cm,
            bb=0 0 500 400,
            draft=false
        ]{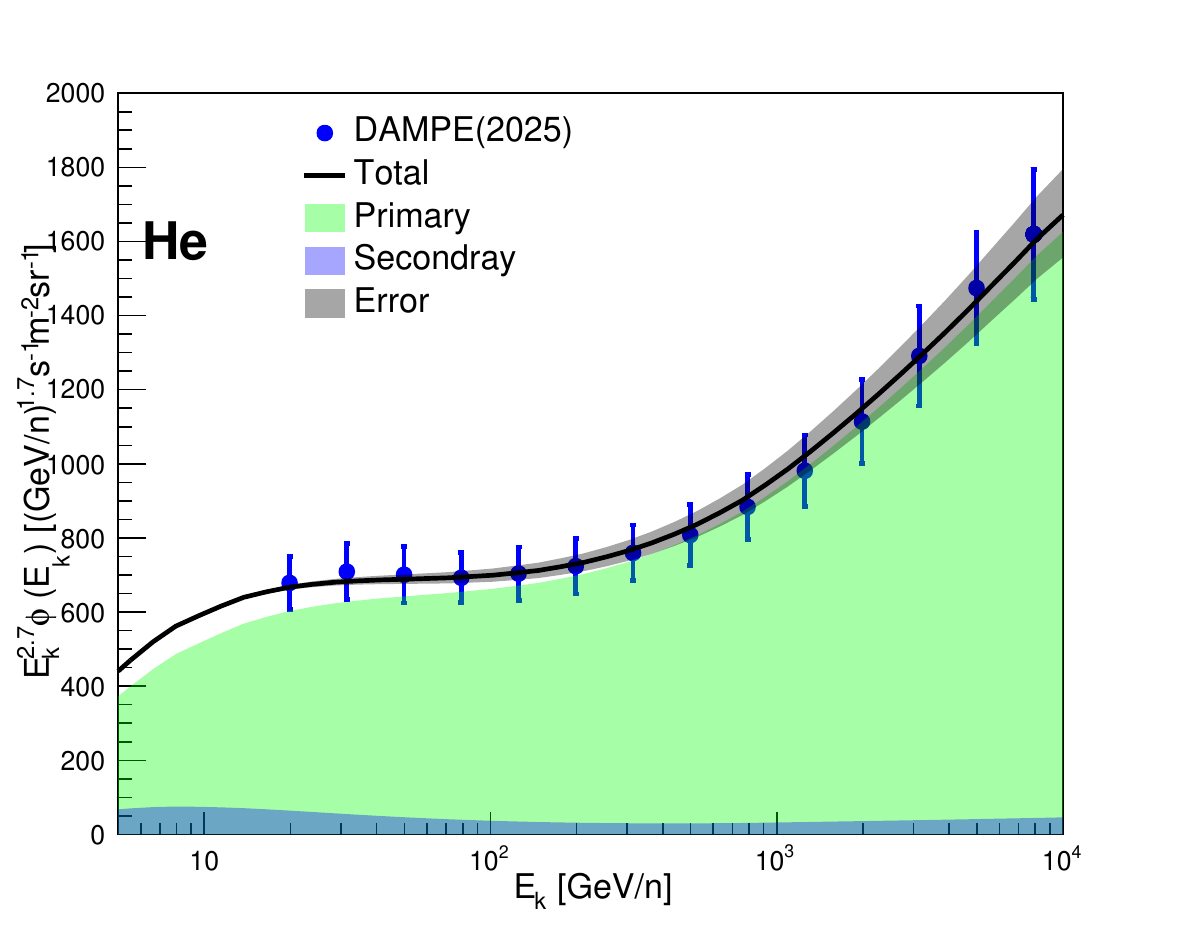}
        \label{fig:he_dampe}
    \end{subfigure}
    \hfill
    \begin{subfigure}[t]{0.24\textwidth}
        \centering
        \includegraphics[
            width=\linewidth,
            height=3.5cm,
            bb=0 0 500 400,
            draft=false
        ]{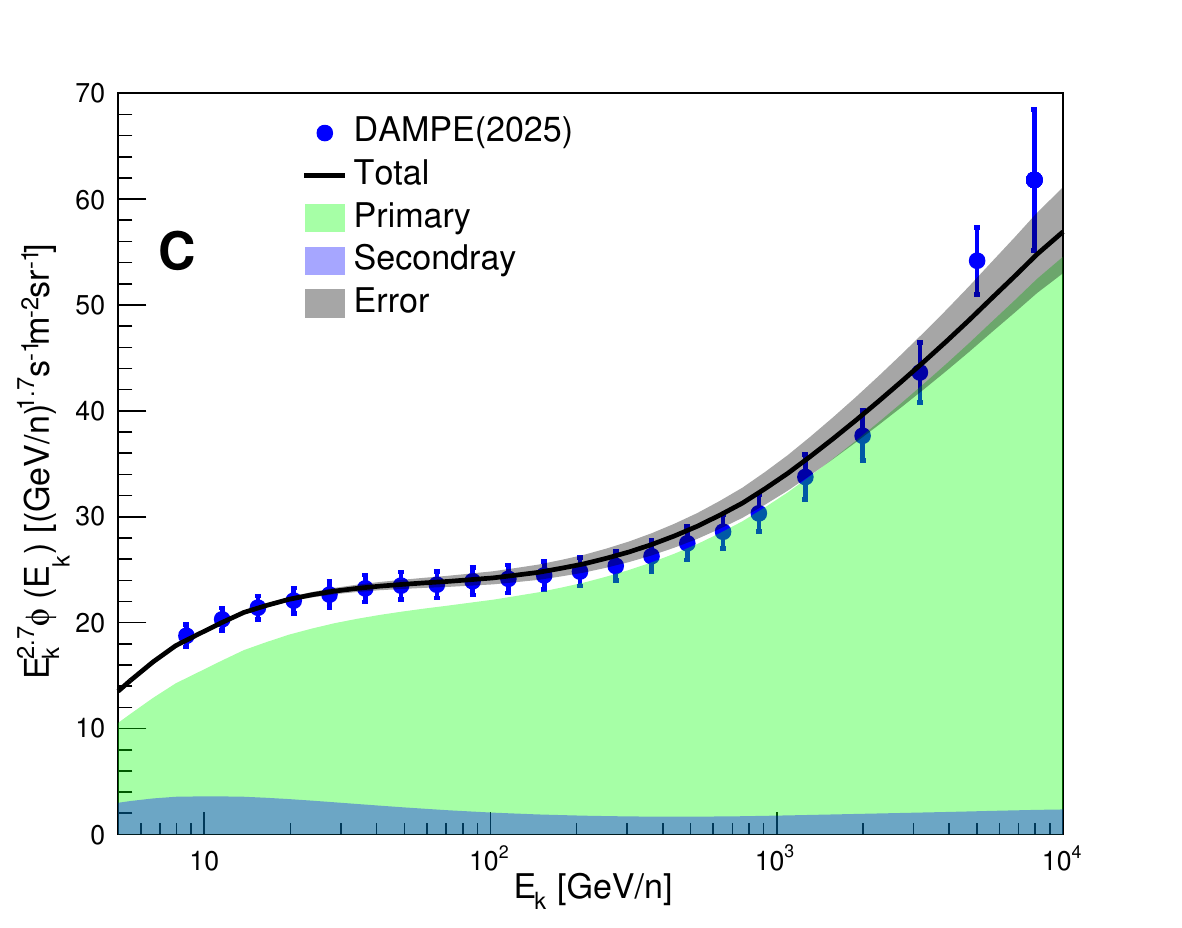}
        \label{fig:c_dampe}
    \end{subfigure}
    \hfill
    \begin{subfigure}[t]{0.24\textwidth}
        \centering
        \includegraphics[
            width=\linewidth,
            height=3.5cm,
            bb=0 0 500 400,
            draft=false
        ]{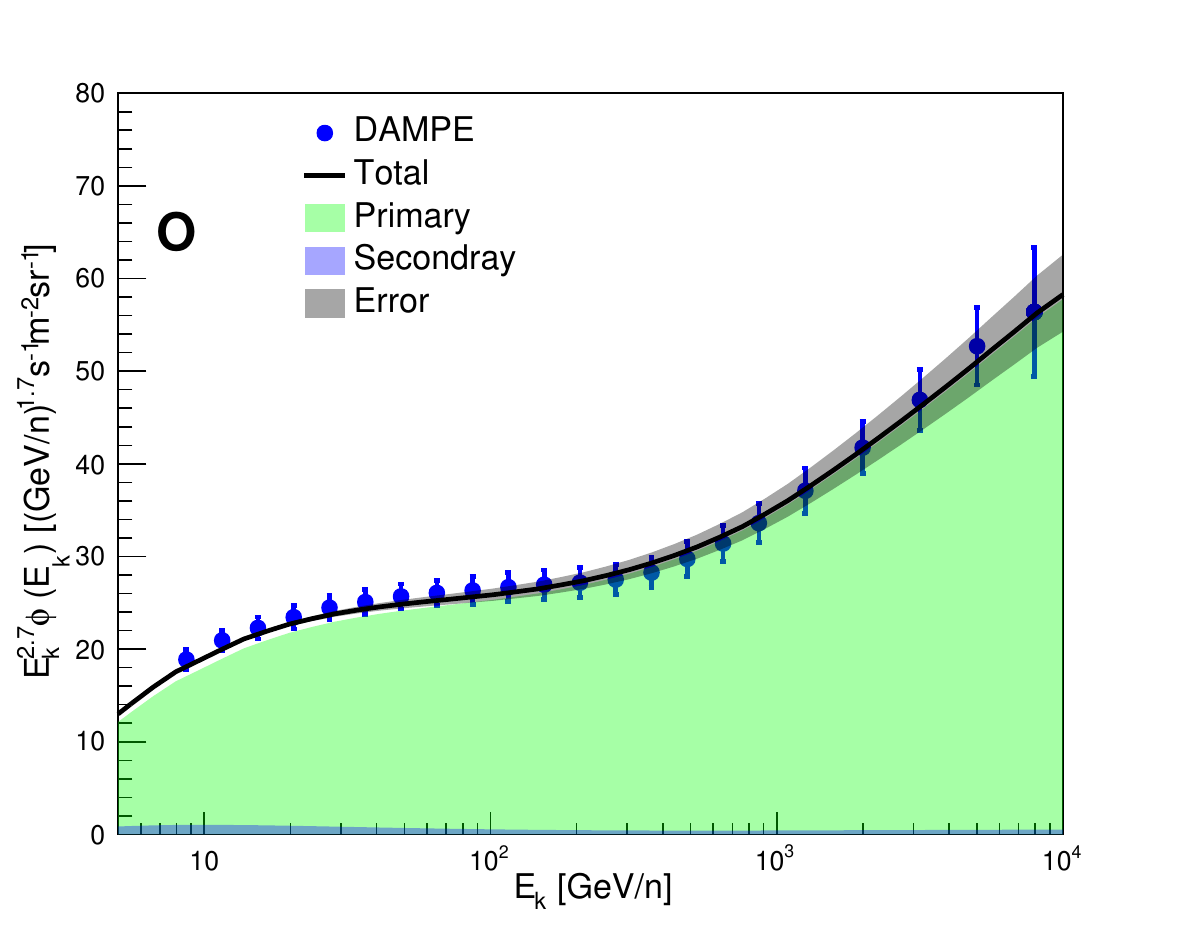}
        \label{fig:o_dampe}
    \end{subfigure}
    \hfill
    \begin{subfigure}[t]{0.24\textwidth}
        \centering
        \includegraphics[
            width=\linewidth,
            height=3.5cm,
            bb=0 0 500 400,
            draft=false
        ]{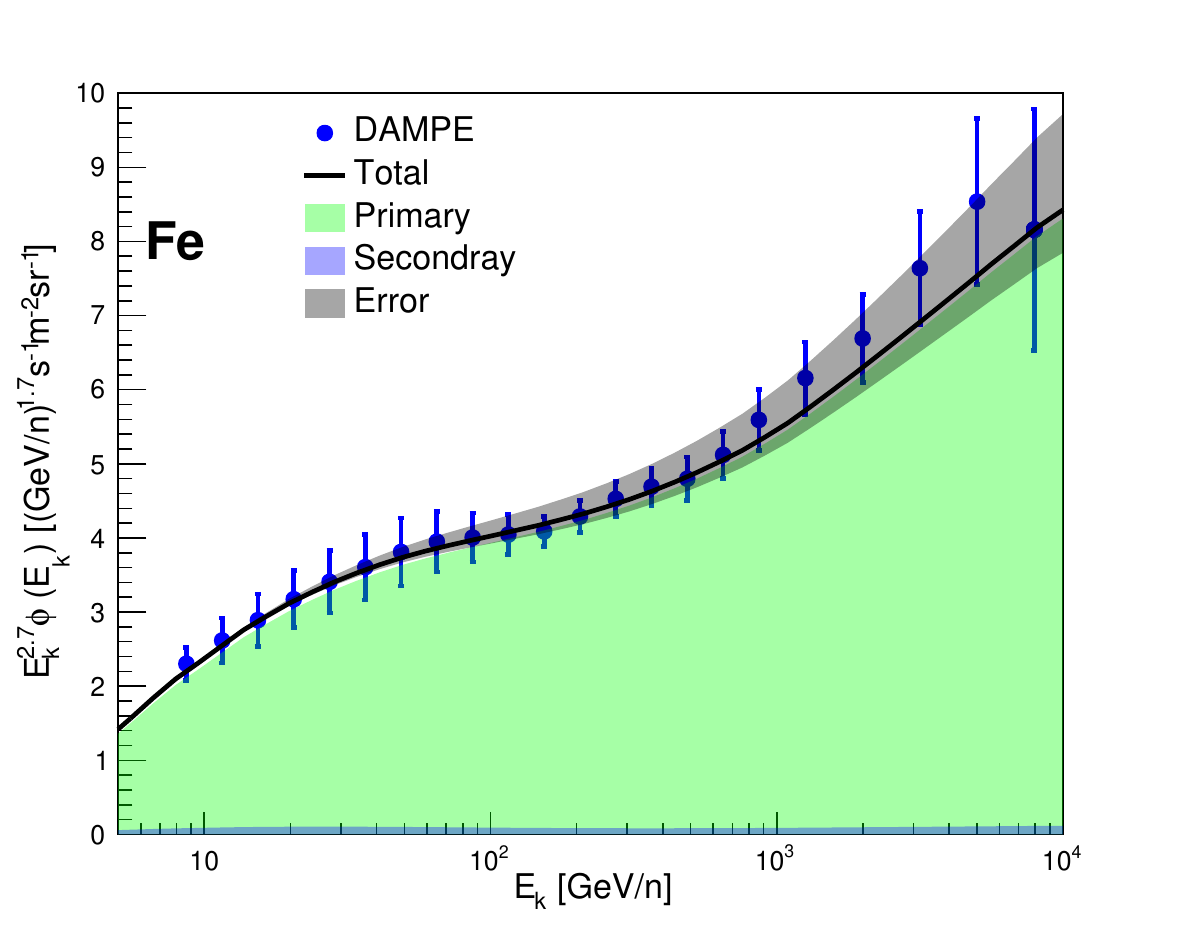}
        \label{fig:fe_dampe}
    \end{subfigure}
    
    \caption{The fluxes of various elements calculated by the model are compared with the experimental observations from AMS-02 and DAMPE.
In the figure, red points represent AMS-02 observational results, blue points represent DAMPE observational results, the black solid line denotes the total flux calculated by the model, and the shaded black area indicates the error range obtained under the condition of $\chi^2/\mathrm{d.o.f.} < 2$. "Primary" and "Secondary" refer to the energy spectra of the primary and secondary components, respectively.   }
    \label{fig:element_spectra_35cm}
\end{figure*}

\bibliography{apssamp}% Produces the bibliography via BibTeX.

\end{document}